\documentclass{article}
\usepackage{spconf}
\usepackage{epsfig}
\usepackage{psfrag}
\usepackage{color}
\usepackage{tabularx}
\usepackage{pstricks, pst-node, pst-plot, pst-circ}
\usepackage{moredefs}

\usepackage{fancyhdr}
\newcolumntype{C}[1]{>{\centering\arraybackslash}p{#1}}

\def\e{\mathrm{ e}}
\def\PSNR{\mathrm{ PSNR}}
\def\punit{\, \mathrm}

\title{Spatio-Temporal Prediction in Video Coding by Non-Local Means Refined Motion Compensation}
\name{J\"urgen~Seiler, Thomas~Richter, and Andr\'e~Kaup}
\address{Chair of Multimedia Communications and Signal Processing, \\University of Erlangen-Nuremberg, Cauerstr. 7, 91058 Erlangen, Germany\\ \{seiler, richter, kaup\}@LNT.de}

\begin{document}
\topmargin=0mm
\maketitle


\begin{abstract} \label{abstract}
The prediction step is a very important part of hybrid video codecs. In this contribution, a novel spatio-temporal prediction algorithm is introduced. For this, the prediction is carried out in two steps. Firstly, a preliminary temporal prediction is conducted by motion compensation. Afterwards, spatial refinement is carried out for incorporating spatial redundancies from already decoded neighboring blocks. Thereby, the spatial refinement is achieved by applying Non-Local Means denoising to the union of the motion compensated block and the already decoded blocks. Including the spatial refinement into H.264/AVC, a rate reduction of up to 14 \% or respectively a gain of up to 0.7 dB PSNR compared to unrefined motion compensated prediction can be achieved.
\end{abstract}




\section{Introduction} \label{sec:introduction} 

Two different strategies can be used for compressing digital signals. The first is redundancy reduction, the second is irrelevance reduction. For digital video signals, both strategies have to be heavily utilized for shrinking the amount of data to a size that can be transmitted or stored. In hybrid video codecs the irrelevance reduction is achieved by quantization, the redundancy reduction is achieved by prediction, transform of the prediction residual and entropy coding. Regarding this chain, one can recognize that, besides the other techniques, the coding efficiency directly depends on the prediction quality. In the prediction step, the area actually being encoded is estimated from already transmitted parts of the sequence, so that the decoder can predict the signal in the same way as the encoder. Thus, only the prediction error has to be transmitted. Obviously, the overall amount of data to be transmitted directly depends on the prediction quality. The better the prediction quality, the less data has to be transmitted and vice versa.

Although typical video signals possess spatial as well as temporal redundancies, modern hybrid video codecs like H.264/AVC only exploit either temporal or spatial ones at a time \cite{Wiegand2003}. In doing so, the temporal prediction is obtained by motion compensation \cite{Dufaux1995}. For spatial prediction, the already transmitted and reconstructed parts of the actually processed frame are skillfully continued into the area to be predicted. Even though H.264/AVC can switch between temporal and spatial prediction, a combined prediction is not used. In literature only very few algorithms for spatio-temporal prediction in video coding exist. Two of them to mention are the Inter Frame Coding with Template Matching Spatio-Temporal Prediction from \cite{Sugimoto2004} or the Joint Predictive Coding from \cite{Jiang2009}. In \cite{Seiler2008c, Seiler2010} we introduced two different algorithms for spatio-temporal prediction, the Spatially Refined Motion Compensation by Frequency Selective Approximation (FSA) or respectively by Multiple Selection Approximation (MSA). These algorithms are able to increase the coding efficiency significantly by exploiting temporal as well as spatial redundancies. For this, first motion compensation is carried out. Afterwards, a joint model of the spatial neighborhood and the motion compensated signal is generated for combining temporal and spatial information. 

In this contribution we introduce the Non-Local Means Refined Prediction, a novel spatio-temporal prediction algorithm based on spatial refinement. This algorithms also makes use of motion compensation for preliminary prediction. \mbox{After} this, a Non-Local Means based denoising algorithm is used for incorporating spatial information from the already decoded signal parts and therewith form a better predictor.


\section{Spatial Refinement by Non-Local Means} \label{sec:nlm-refinement}

\begin{figure}
	\begin{center}
		\includegraphics[width=0.45\textwidth]{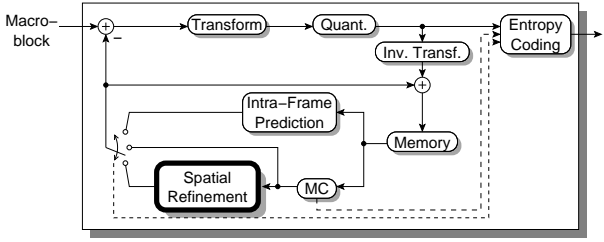}
	\end{center}\vspace{-0.4cm}
	\caption{Block diagram of a hybrid video encoder with spatial refinement.\vspace{-0.1cm}}
	\label{fig:coder_diagram}
\end{figure}

The Non-Local Means Refined Prediction (NLM-RP) is carried out in two steps. The first step is to conduct motion compensated prediction for the block to be encoded, as usual in most hybrid video codecs. Using this, a good initial estimate of the block to be predicted can be obtained, only at the expense of the motion vector that has to be transmitted as side information. The second step is the subsequent outlined spatial refinement for exploiting the spatial redundancies to the already decoded neighboring blocks. Fig.\ \ref{fig:coder_diagram} shows a simplified block diagram of a hybrid video codec that makes use of spatio-temporal prediction by spatially refined motion compensation. 

For describing the spatial refinement, we regard the video sequence $v\left[x,y,t\right]$ with spatial coordinates $x,y$ and temporal coordinate $t$. The macroblock actually being coded is at position $\left(x_0,y_0\right)$ in frame $t=\tau$ and the sequence is processed in line scan order. Thus, all macroblocks left and above the actually processed one have been transmitted and decoded. As they are available at the decoder as well, they can be used for prediction. A preliminary prediction for the macroblock actually being processed is obtained by motion compensation from one of the previous frames. By transmitting the motion vector as side information, the decoder can obtain the same preliminary estimate by selecting the corresponding area from the previous frame.

\begin{figure}
	\psfrag{m}[t][t][0.9]{$m$}%
	\psfrag{n}[t][t][0.9]{$n$}%
	\psfrag{x}[t][t][0.9]{$x$}%
	\psfrag{y}[t][t][0.9]{$y$}%
	\psfrag{t}[t][t][0.9]{$t$}%
	\psfrag{B}[l][l][0.9]{$\mathcal{B}$}%
	\psfrag{R}[l][l][0.9]{$\mathcal{R}$}%
	\psfrag{x0}[t][t][0.9]{$x_0$}%
	\psfrag{y0}[t][t][0.9]{$y_0$}%
	\psfrag{tau}[t][t][0.9]{$\tau$}%
	\psfrag{tau1}[t][t][0.9]{$\tau-1$}%
	\centering
	\includegraphics[width=0.45\textwidth]{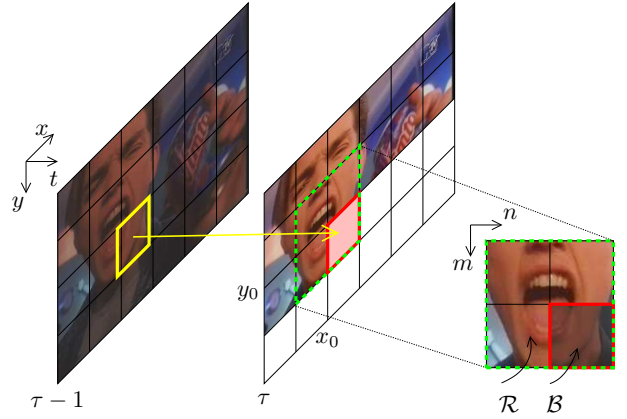}
	\caption{Relation between video sequence and processing area $\mathcal{L}$ as union of reconstructed blocks $\mathcal{R}$ and motion compensated block $\mathcal{B}$.\vspace{-0.1cm}}
	\label{fig:refinement_area}
\end{figure}

To achieve spatial refinement, an area of four macroblocks is cut out of the sequence. This so called processing area $\mathcal{L}$ consists of the actual block to be predicted and the three already decoded macroblocks to the left and above. As mentioned before, for the block to be predicted a preliminary prediction is obtained by motion compensation. The relation between the video sequence and area $\mathcal{L}$ is shown in Fig.\ \ref{fig:refinement_area}. The signal $\widetilde{s}\left[m,n\right]$ in area $\mathcal{L}$ is indicated by the two spatial coordinates $m$ and $n$. Area $\mathcal{L}$ can be split in two sub-areas. On the one hand, this is  area $\mathcal{R}$ which contains all already reconstructed blocks. On the other hand, area $\mathcal{B}$ subsumes all samples that result from the preliminary motion compensation from a previous frame. Fig.\ \ref{fig:refinement_area} also shows the relationship between the different areas. 

Since the already decoded blocks in $\mathcal{R}$ suffer from quantization, there the signal
\begin{equation}
	\widetilde{s}\left[m,n\right] = s\left[m,n\right] + n_\mathrm{q}\left[m,n\right], \forall \left(m,n\right)\in\mathcal{R}
\end{equation} 
could be regarded as the original, unquantized signal $s\left[m,n\right]$ superimposed by the quantization error $n_\mathrm{q}\left[m,n\right]$. Regarding the preliminary predicted signal in area $\mathcal{B}$, there the signal  
\begin{equation}
	\widetilde{s}\left[m,n\right] = s\left[m,n\right]+ n_\mathrm{q}\left[m,n\right] + n_\mathrm{p}\left[m,n\right], \forall \left(m,n\right)\in\mathcal{B}
\end{equation} 
that results from motion compensation is the original signal $s\left[m,n\right]$ superimposed by the prediction error $n_\mathrm{p}\left[m,n\right]$ and also the quantization error $n_\mathrm{q}\left[m,n\right]$. The quantization error emerges as the preliminary motion compensated prediction is based on the quantized data in the reference frame. In addition to that, all the parts of the signal that cannot be covered by motion compensation are subsumed in $n_\mathrm{p}\left[m,n\right]$.

Obviously, the coding efficiency directly depends on the ability of a codec to predict the signal based on the already transmitted areas. Hence, the aim is to estimate $s\left[m,n\right]$ in area $\mathcal{B}$ in the best possible way from the available signal $\widetilde{s}\left[m,n\right]$. Regarding the equations above, one can recognize that the distorting terms have to be suppressed in order to retrieve the original signal $s\left[m,n\right]$ from $\widetilde{s}\left[m,n\right]$. To achieve this, we regard the quantization error $n_\mathrm{q}\left[m,n\right]$ and the prediction error $n_\mathrm{p}\left[m,n\right]$ as noise terms and propose to use the Non-Local Means (NLM) denoising algorithm from \cite{Buades2005} for reducing them. NLM is well suited for this purpose, as it exploits self similarities within a signal for reducing the distortion. Due to the non-local character of NLM, the spatial correlations within a video signal can be exploited in addition to the temporal ones which have already been exploited by motion compensation. 

For refinement, NLM is applied to the union of the reconstructed blocks in $\mathcal{R}$ and the motion compensated block in $\mathcal{B}$. As only region $\mathcal{B}$ is used for the final prediction, the refined signal $\hat{s}\left[m,n\right]$ has to be determined only in this area. According to \cite{Buades2005}, the refined values result from the weighted average of all pixels in the image. With that, the refined samples are given by
\begin{equation}
\label{eq:refinement}
 \hat{s}\left[m,n\right]=\frac{\displaystyle\sum_{\left(k,l\right)\in\mathcal{L}} \widetilde{s}\left[k,l\right] w_{\left(m,n\right)}\left[k,l\right]}{\displaystyle\sum_{\left(k,l\right)\in\mathcal{L}} w_{\left(m,n\right)}\left[k,l\right]},\ \forall \left(m,n\right)\in\mathcal{B}.
\end{equation}
In order to keep the refinement step manageable, in our case not the complete area of the frame that already has been decoded is used, but only the three neighboring already decoded macroblocks. The weight which is assigned to every pixel during the averaging depends on its own postion and on the position of the sample to be refined. The summation over the individual weights that is carried out in the denominator is required only for normalization. For calculating the individual weights of every sample, the sum of squared differences
\begin{equation}
\label{eq:distance}
 d_{\left(m,n\right)}\left[k,l\right] = \hspace{-3mm}\sum_{{\mu, \nu=}\atop{-d_\mathrm{m},\ldots, d_\mathrm{m}}}\hspace{-3mm} \left(\widetilde{s}\left[m+\mu,n+\nu\right]-\widetilde{s}\left[k+\mu,l+\nu\right]\right)^2
\end{equation} 
between two neighborhoods around the pixel $\widetilde{s}\left[m,n\right]$ to be refined and around the candidate pixel $\widetilde{s}\left[k,l\right]$ is regarded. Each of the neighborhoods has a width of $2d_\mathrm{m}+1$ samples. Based on this, the actual weights result from
\begin{figure}
	\psfrag{sm}[l][l][0.9]{$\widetilde{s}\left[m,n\right]$}%
	\psfrag{sk1}[l][l][0.9]{$\widetilde{s}\left[k_1,l_1\right]$}%
	\psfrag{sk2}[l][l][0.9]{$\widetilde{s}\left[k_2,l_2\right]$}%
	\psfrag{sk3}[l][l][0.9]{$\widetilde{s}\left[k_3,l_3\right]$}%
	\centering
	\includegraphics[width=0.2\textwidth]{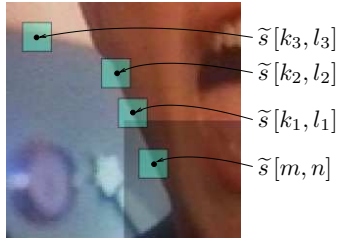}\vspace{-0.1cm}
	\caption{Spatial refinement of pixel $\widetilde{s}\left[m,n\right]$. Due to the similar neighborhoods, samples like $\widetilde{s}\left[k_1,l_1\right]$ and $\widetilde{s}\left[k_2,l_2\right]$ get high weights, whereas ones like $\widetilde{s}\left[k_3,l_3\right]$ only obtain small weights.\vspace{-0.1cm}}
	\label{fig:nlm_refinement}
\end{figure}
\begin{equation}
 \label{eq:weight}
 w_{\left(m,n\right)}\left[k,l\right] = \e^{-d_{\left(m,n\right)}\left[k,l\right] / h^2}.
\end{equation}
Here, in addition to the sum of squared differences, the factor $1/h^2$ is included in the exponent for controlling the behavior of the weight assignment. Regarding (\ref{eq:distance}) and (\ref{eq:weight}) in detail, one can recognize that the weight of a candidate pixel in the summation directly depends on the similarity between the neighborhood surrounding this pixel and the neighborhood surrounding the pixel to be refined. The more similar the neighborhoods are, the higher is the weight and the sample gets more influence. On the other hand, if the neighborhoods surrounding the samples differ, the weight gets small and the influence of this particular sample decreases. Due to the position of the sum of squared differences in the exponent, the weight changes very fast for different values of $d_{\left(m,n\right)}\left[k,l\right]$. In addition to that, parameter $h$ controls the amount of averaging. If $h$ is set to a large value, even samples that lie in a very distinct neighborhood obtain a relatively high weight in (\ref{eq:refinement}). For small values of $h$, only neighborhoods that are very similar to the one around $\widetilde{s}\left[m,n\right]$ can result in significant weights. With this calculation of the individual weights, spatial information from the already reconstructed macroblocks can be drawn into the motion compensated block. In Fig.\ \ref{fig:nlm_refinement} the refinement is shown exemplarily for one pixel. Thereby, samples like $\widetilde{s}\left[k_1,l_1\right]$ or $\widetilde{s}\left[k_2,l_2\right]$ get a high weight during the averaging, as their neighborhoods are similar to the neighborhood around $\widetilde{s}\left[m,n\right]$. Samples like $\widetilde{s}\left[k_3,l_3\right]$ on the other hand side only get a small weight, as their surrounding neighborhood is unlike the neighborhood around $\widetilde{s}\left[m,n\right]$. In the case that a sample is located at the border of $\mathcal{B}$, the regarded neighborhoods are not square anymore but have to be shrunk to a size so that no samples lying outside $\mathcal{B}$ are referenced.

After the refinement has been carried out for every sample in $\mathcal{B}$, the refined signal $\hat{s}\left[m,n\right]$ is used for prediction. Since the already reconstructed blocks are used for reference in addition to the motion compensated block, temporal as well as spatial redundancies are exploited at the same time and a higher prediction quality can be achieved.


\section{Simulation Setup and Results} \label{sec:results}

For evaluating the abilities of NLM-RP, the novel refinement is included in the H.264/AVC reference software JM10.2 \cite{JVT2007} with Baseline Profile and Level 2.0 . The motion compensation which provides the preliminary prediction is carried out with quarter pixel accuracy and a maximum search range of $16$ samples. In order to compare NLM-RP to the original unrefined motion compensated prediction and to spatial refinement by FSA \cite{Seiler2008c} and MSA \cite{Seiler2010} the CIF sequences ``Crew'', ``Foreman'' and ``Vimto'' are encoded at $30$ frames per second. Thereby, the rate control is switched off and $10$ different fixed QPs from the range between $16$ and $43$ are examined.

The extent of the neighborhoods is set to $d_\mathrm{m}=3$ and the parameter $h$ is set to $25$. Fortunately, none of the two parameters is very critical and both can be varied widely without significantly affecting the prediction quality. For sequences where motion compensation often leads to inappropriate prediction, larger values of $h$ would improve the prediction quality as the amount of averaging is increased. On the other hand, a too strong averaging would decrease the quality of the refinement for sequences where motion compensation already leads to a good initial estimate. The values listed above provide a good trade-off and lead to a good refinement for all tested sequences. Furthermore, it has to be noted that the spatial refinement does not necessarily lead to a better predictor than the underlying motion compensated prediction. To account for this, one bit of side information has to be transmitted per macroblock to signalize the decoder if the spatial refinement has to be used or not. 

\begin{figure}
	\centering
	\input{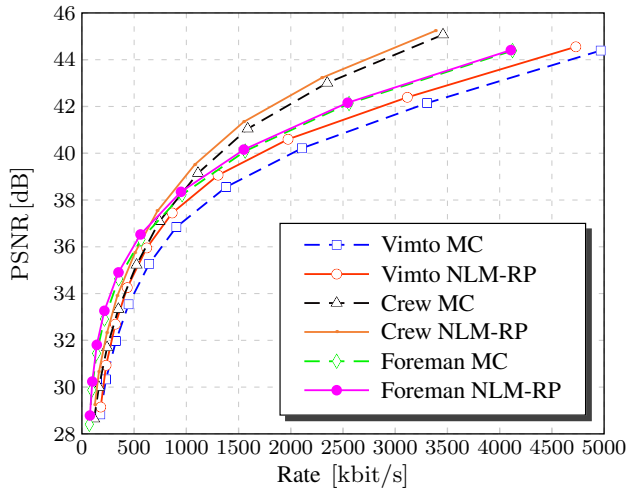}\vspace{-0.1cm}
	\caption{Rate-distortion curves for first $99$ P-frames. Comparison between unrefined motion compensated prediction and Non-Local Means Refined Prediction.\vspace{-0.1cm}}
	\label{fig:rd_plot}
\end{figure}

Fig.\ \ref{fig:rd_plot} shows the rate distortion curves for the aforementioned sequences. The prediction is carried out either by motion compensation without any refinement (MC) or with refinement carried out by NLM as described above. Obviously, over the complete quality range a significant gain can be achieved by spatial refinement of the motion compensated predictor. Thereby, the gain is larger for sequences like ``Crew'' or ``Vimto'' as motion compensation often fails there due to non-translational motion or illumination changes. For sequences like ``Foreman'', where motion compensation forms an almost perfect predictor in most cases, of course the gain achievable by spatial refinement is smaller.

In order to quantify the gain that can be achieved for every sequence, the mean gain over the complete quality range is calculated according to the Bj{\o}ntegaard metric \cite{Bjontegaard2001}. Table \ref{tab:results} lists the mean rate reduction or respectively the mean $\PSNR$ gain produced by spatial refinement compared to unrefined motion compensated prediction. In addition to \mbox{NLM-RP}, the two earlier refinement algorithms FSA \cite{Seiler2008c} and MSA \cite{Seiler2010} are listed for comparison. It can be recognized that independently of the actual refinement algorithm the data rate can be reduced by up to $14\%$ or a gain of up to $0.7 \punit{dB}$ could be achieved, respectively. Considering all the regarded sequences, a mean rate reduction of $9.6 \%$ or a mean $\PSNR$ gain of $0.47 \punit{dB}$ is feasible. Comparing the refinement algorithms among each other, one can further recognize that NLM-RP is able to even outperform MSA by $1.3$ percentage points in rate reduction. Thereby it has to be considered that MSA already was able to achieve a significant gain over MC.


\section{Conclusion} \label{sec:conclusion}

In this paper, a novel algorithm for spatio-temporal prediction in video coding was introduced. This algorithm works on top of a preliminary motion compensated prediction. The initial temporally extrapolated signal is spatially refined in order to exploit spatial redundancies in addition to the already exploited temporal ones. For this purpose, the spatial refinement is conducted by applying Non-Local Means denoising to the union of the already reconstructed signal parts and the preliminary motion compensated block. With this novel algorithm a mean rate reduction of $9.6 \%$ or respectively a mean $\PSNR$ gain of $0.47 \punit{dB}$ is achieved compared to unrefined motion compensated prediction.

Although the coding efficiency already can be significantly increased by the proposed spatio-temporal prediction algorithm, alternative denoising algorithms will be evaluated as well in the future, in order to determine their abilities for spatial refinement. In addition to that, current research aims at incorporating the statistics of the quantization and the prediction error into the denoising process for further improving the spatial refinement.

\begin{table}
\small
\begin{tabular}{l|c|c|c||c}
& ``Crew'' & ``Foreman'' & ``Vimto'' & Mean\\ \hline \hline
\multicolumn{4}{l}{Rate reduction}\\ \hline
FSA \cite{Seiler2008c} & $7.32\%$ & $3.20\%$ & $13.42\%$ & $7.98\%$ \\ \hline
MSA\cite{Seiler2010} & $7.69\%$ & $2.26\%$ &  $14.98\%$ & $8.31\%$ \\ \hline
NLM-RP & $10.44\%$ & $4.77\%$ & $13.68\%$ & $9.63\%$ \\ \hline
\multicolumn{4}{l}{$\PSNR$ gain}\\ \hline
FSA \cite{Seiler2008c} & $0.37\punit{dB}$ & $0.13\punit{dB}$ & $0.66\punit{dB}$ & $0.39\punit{dB}$ \\ \hline
MSA\cite{Seiler2010} & $0.39\punit{dB}$ & $0.09\punit{dB}$ & $0.74\punit{dB}$ & $0.41\punit{dB}$  \\ \hline
NLM-RP & $0.54\punit{dB}$ & $0.19\punit{dB}$ & $0.67\punit{dB}$ & $0.47\punit{dB}$ 
\end{tabular}\vspace{-0.1cm}
\caption{Rate reduction and $\PSNR$ gain compared to pure motion compensation, calculated according to \cite{Bjontegaard2001}.\vspace{-0.1cm}}
\label{tab:results}
\end{table}




\end{document}